\documentclass[11pt,english]{article}
\usepackage[T1]{fontenc}
\usepackage[english]{babel}

\usepackage[top=1in,bottom=1in,left=0.75in,right=0.75in]{geometry}
\usepackage{times}
\usepackage{amsmath,amssymb,amsfonts}
\usepackage{graphicx}

\usepackage{natbib}
\usepackage{booktabs}
\usepackage{caption}
\usepackage{subcaption}
\usepackage{float}

\usepackage{hyperref}
\hypersetup{
  colorlinks   = true,
  linkcolor    = blue,
  citecolor    = blue,
  urlcolor     = cyan
}

\usepackage{titlesec}
\titleformat{\section}
  {\normalfont\large\bfseries}{\thesection.}{1em}{}
\titleformat{\subsection}
  {\normalfont\normalsize\bfseries}{\thesubsection}{1em}{}
\titlespacing*{\section}{0pt}{12pt}{6pt}
\titlespacing*{\subsection}{0pt}{12pt}{6pt}

\usepackage[flushmargin]{footmisc}

\usepackage[stretch=10,shrink=10,step=1]{microtype}
\usepackage{enumitem}

\usepackage{tabularx}
\newcolumntype{L}{>{\raggedright\arraybackslash}X}
\begin{document}
\twocolumn[
  \begin{center}
    {\LARGE Technical Overview of Safe3Step (S3S): Power Ratings and quality wins for selecting at-large teams to the NCAA Division I Men's Lacrosse Championship \par}
    \vskip 1.5em
    {\large
      \lineskip 2em
      \begin{tabular}[t]{c}
        Lawrence Feldman$^{1}$ \\
        \texttt{laf@laxnumbers.com}
        \and
        Matthew Bomparola$^{2}$ \\
        \texttt{matthewbomparola@gmail.com}
      \end{tabular}\par}
    \vskip 1.5em
    Contributors:\\\textit{The Men's Division I Lacrosse Selection Criteria and Ranking Committee (SCR),\\ chaired by UVA Head Coach Lars Tiffany$^{3}$.}
  \end{center}
  
  \vskip 1.5em
  \noindent\textbf{Abstract:} This document describes a system for selecting teams to the NCAA Men's Division I Lacrosse Championship Tournament called "Safe3Step" (S3S) that was developed in collaboration with the NCAA Lacrosse Selection Criteria and Ranking Committee (SCR) with the objective of improving on the Ratings Percentage Index (RPI). S3S employs three steps that: 1) evaluate the strength of each team based on score data, 2) award S3S points to each team based on the quality of its wins and losses, ranking teams accordingly and 3) examine each pair of teams with adjacent rankings, swapping ranks if the lower-ranked team has a better head-to-head record against the higher-ranked team. Safe3Step is not entirely new, but it improves on other "quality win" methods by using Power Ratings to identify team strengths, respecting head-to-head records, and adhering to standards of simplicity, transparency, and objectivity. Empirical analysis is left to future work.
  
  \vskip 0.5em
  \noindent\textbf{Keywords:} At-large selection; Division I Lacrosse; quality wins; Power Ratings; goal differences; Powerwise; RPI. 
  
  \vskip 2em
]

\footnotetext[1]{Lawrence Feldman played lacrosse at the University of Pennsylvania, received a PhD from Auburn University and attended Stanford University as a Post Doc. He worked as an engineer/computer analyst in the USAF and worked at Lawrence Livermore Laboratory, Los Alamos National Laboratory, NASA, and Intel Corp. Feldman has served on the Coaches Committee for Division I men's lacrosse for the past two years as an advisor. He founded Laxpower, a web site that rated lacrosse teams from 1997 to 2019.}

\footnotetext[2]{Matthew Bomparola has played and coached tennis for two decades. He received a degree from Princeton University in '21 and works as a data analyst, writer, and journalist.}

\footnotetext[3]{The SCR includes: Andrew Shay, Yale; William Wilson, Air Force Academy; Ryan Danehy, Mercer; J.L. Reppert, Holy Cross; Matthew Madalon, Princeton; Keegan Wilkinson, Marist; Marc Van Arsdale, Loyola; Chris Wojcik, Notre Dame; Lars Tiffany (Chairman), Virginia; Kevin Conry, Michigan; Doug Murray, Air Force Academy.}

\section{Introduction}
Safe3Step (S3S) is one of a handful of methods developed by Lawrence Feldman with input from the Lacrosse Selection Criteria and Ranking Committee (abbreviated SCR, or simply the “committee”) with the goal of creating a ranking system uniquely suited to selecting teams to the NCAA Men’s Division I Lacrosse Championship Tournament.\footnote{Another method developed in collaboration with the Selection Criteria and Ranking Committee is called Powerwise (PWR). Its pairwise methodology also makes use of Power Ratings and is described in \cite{FeldmanBomparola2025}.} In four years, the committee examined more than a dozen potential methods, testing each against 20 years of historical lacrosse data to compare its hypothetical at-large selections with real historical selections.

The existing selection process—involving closed-door committees informed by experts and the flawed Ratings Percentage Index (RPI) for statistical support\footnote{See Appendix A for examples of the RPI's flaws.}—produces reasonable results, but its lack of transparency has led to, in the words of academics writing about another NCAA sport, a “large amount of speculation, second guessing, and debate each year about decisions.” \cite{Coleman2010} S3S was found to be simpler, more objective, and more transparent than the existing method, meeting the following criteria identified as desirable by Feldman and the committee: \\

\newcounter{myroman}
\begin{list}{\Roman{myroman}.}{%
    \usecounter{myroman}%
    \setlength{\leftmargin}{1.5em}
    \setlength{\labelwidth}{1.5em}%
    \setlength{\labelsep}{0.5em}
    \setlength{\itemsep}{2pt}
    \setlength{\topsep}{0pt}
    \setlength{\parsep}{0pt}%
    \setlength{\partopsep}{0pt}%
}
  \item Producing consistent, common-sense, and easily reproducible results.
  \item Preferring simple data like head-to-head results to complex analytics.
  \item Using all interpretable results data, including game scores.
  \item Adjusting for strength of schedule without incentivizing gamesmanship.
  \item Minimizing bias and human intervention.\\
\end{list}

This overview of Safe3Step describes the three-step ranking procedure and provides a small example calculation. Empirical evaluation and analysis are left to future work.

\section{The Safe3Step Method}
In short: S3S 1) analyzes score data to determine team strength, thereby 2) assigning points to teams based on the quality of their regular-season results before 3) flipping any adjacent rankings in which the lower-ranked team beat the higher-ranked team on the field. 

\subsection*{Step I: Assessing team strength}
S3S uses a Power Rating (PR) algorithm to determine team strength. The formula is a bit mathematical, but it's simpler than it looks:\footnote{Standard summation notation indicates an iterative process in which all matchups are considered between any two teams \textit{i} and \textit{j}. $PR_i - PR_j$ = difference in estimated Power Ratings; $score_i - score_j$ = real (measured) difference in scores; $hfa$ = home-field advantage.}

\begin{equation}\label{eq:power-rating}
  \sum_{i=0}^n \sum_{j=0}^m (PR_i - PR_j)
    = \sum_{i=0}^n \sum_{j=0}^m (score_i - score_j)
    \pm hfa
\end{equation}

Calculating Power Ratings is an iterative process that solves a large system of linear equations with variables for real game scores, power ratings, and the home-field advantage.\footnote{Strength of schedule is automatically accounted for thanks to the inclusion of score data. The home-field advantage is a constant 0.73 goals for the away team, but it can float or be adjusted as needed.} The algorithm iterates until the average difference between each pair of teams' Power Ratings is equal to the expected (or real, if the teams really played) difference in scores were those teams to face each other on a neutral field. 

Once Power Ratings are computed, they're adjusted to create a point allocation table that determines how many S3S points are awarded for regular season wins and losses against any team in the division. 

\begin{table}[H]
\centering
\begingroup
\footnotesize
\setlength{\tabcolsep}{6pt}
\caption{Power Ratings and S3S Points \\(Men's Division I Lacrosse, 2024–2025)}
\label{tab:s3s-unified-top10}
\begin{tabular*}{\linewidth}{@{\extracolsep{\fill}} c l c c c c @{}}
\toprule
Rank & Team & Record & PR & 99.9\,$-$\,PR* & \(|\cdot-25|\)** \\
\midrule
1  & Notre Dame   & 10--2 & 99.90 & 0.00 & 25.00 \\
2  & Virginia     & 11--3 & 99.73 & 0.17 & 24.83 \\
3  & Duke         & 13--2 & 97.89 & 2.01 & 22.99 \\
4  & Penn State   &  9--4 & 97.16 & 2.74 & 22.26 \\
5  & Maryland     & 10--5 & 97.12 & 2.78 & 22.22 \\
6  & Cornell      & 11--3 & 96.81 & 3.09 & 21.91 \\
7  & Georgetown   & 12--3 & 96.68 & 3.22 & 21.73 \\
8  & Michigan     &  9--6 & 96.61 & 3.29 & 21.71 \\
9  & Yale         &  9--5 & 96.43 & 3.47 & 21.53 \\
10 & Princeton    &  8--6 & 96.24 & 3.65 & 21.34 \\
\bottomrule
\end{tabular*}
\endgroup
\\
\raggedright\scriptsize
\vspace{2pt}
* S3S points deducted in a loss to this team. \\*** S3S points gained by defeating this team. · is a placeholder for 99.9 — PR.
\end{table}

Table 1 is an extract of the point allocation table for the Men's Division I 2024–2025 lacrosse season. The $99.9-\text{PR}$ column indicates S3S points lost in a defeat against the team in the same row of the table, and the $|99.9-\text{PR}-25|$ column corresponds to points gained in a victory. 

For instance, a loss against Notre Dame (the top team) results in a 0-point deduction to a team's overall S3S score. Conversely, a win against the Fighting Irish results in a 25-point increase, ignoring adjustments for the home-field advantage.\footnote{The number 25 was chosen to equal the approximate difference in PR between the highest and lowest rated teams in the dataset. This constant may be adjusted seasonally or at committee discretion.}

\subsection*{Step II: Assigning S3S points}

Step II involves tallying up S3S points for each game played by each team in the ranking list. Table 2 displays an example tally for Virginia's 2024–2025 season. Note that, even though Virginia only played 14 games, it's S3S score is normalized to 16 games for comparison purposes.\footnote{ \textit{Normalized score = }$(S3S.score/Games.played)*16$, 16 being the number of regular-season games played in a typical season.}

\begin{table}[H]
\centering
\begingroup
\scriptsize
\setlength{\tabcolsep}{5pt}
\caption{Safe3Step Points Tally\\(Virginia's Men’s Division I Lacrosse 2024–2025)}
\label{tab:s3s-season-tally}
\begin{tabular*}{\linewidth}{@{\extracolsep{\fill}} c l c c c c @{}}
\toprule
G\# & Opponent & Score & W/L Pts & HFA & S3S Pts \\
\midrule
1  & Michigan      & 17--13 & +21.71 & $-0.73$ & 20.98 \\
2  & Harvard       & 25--21 & +19.04 & $-0.73$ & 18.31 \\
3  & Ohio St.      & 17--6  & +19.59 & $-0.73$ & 18.85 \\
4  & Richmond      & 25--8  & +19.82 & $-0.73$ & 19.09 \\
$\vdots$ &  &  &  &  & $\vdots$ \\
11 & Duke          & 14--15 & $-2.01$ & $+0.73$ & $-1.28$ \\
12 & Syracuse      & 19--12 & +20.57 & $-0.73$ & 19.83 \\
13 & Lafayette     & 20--11 & +16.57 & $+0.73$ & 17.31 \\
14 & Notre Dame    & 12--8  & +25.00 & $-0.73$ & 24.27 \\
\midrule
\multicolumn{5}{r}{\textit{Season total (raw):}} & \textbf{217.69} \\
\multicolumn{5}{r}{\textit{Normalized to 16 games:}} & \textbf{248.79} \\
\bottomrule
\end{tabular*}
\endgroup
\end{table}

\subsection*{Step III: Swapping head-to-head discrepancies}
Finally, teams are ranked by their ordered normalized S3S point totals. A final pass is made starting from the top of the ranking list, wherein each pair of adjacently-ranked teams is assessed to ensure that the higher ranked team has a winning or tied record against the lower-ranked team. If the lower-ranked team has the better record, their ranks are swapped.\footnote{Only one pass is made.} Table 3 displays example results.\footnote{Coincidentally, no swaps were made for the teams displayed in the 2024–2025 season.} 

\begin{table}[H]
\centering
\begingroup
\small
\setlength{\tabcolsep}{4pt}%
\renewcommand{\arraystretch}{1.08}%
\caption{S3S Ranking List \\(Men’s Division I Lacrosse 2025)}
\label{tab:s3s-ranking-top5-bottom2}
\begin{tabular}{@{} c l c c @{}}
\toprule
Rank & Team & Record & S3S Points \\
\midrule
1  & Notre Dame   & 10--2 & 272.27 \\
2  & Duke         & 13--2 & 269.68 \\
3  & Virginia     & 11--3 & 248.79 \\
4  & Cornell      & 11--3 & 229.13 \\
5  & Georgetown   & 12--3 & 228.08 \\
$\vdots$ &  &  & $\vdots$ \\
19 & Jacksonville & 12--4 & 152.04 \\
20 & Lehigh       & 10--5 & 151.39 \\
\bottomrule
\end{tabular}
\endgroup
\end{table}

\section{Example S3S-ready Dataset}

Safe3Step runs on remarkably simple data—given constants and formulas, all one needs to run S3S is a dataset of every game played by all teams in a season with entries for a game identifier, the teams that played, the scoreline, and an identifier for which team held the home-field advantage (if any\footnote{Neutral-field games are rare, but the edge case should be easy to handle.}).

\begin{table}[H]
\centering
\begingroup
\small
\setlength{\tabcolsep}{4.5pt}%
\renewcommand{\arraystretch}{1.08}%
\caption{Example S3S-ready Dataset\\(Continue for all games in season)}
\label{tab:s3s-input-example}
\begin{tabular}{@{} c l l c c @{}}
\toprule
Game \# & Team 1 & Team 2 & Score & Home Team \\
\midrule
211 & Virginia & Michigan   & 17--13 & Virginia \\
218 & Brown & Harvard    & 25--21 & Brown \\
225 & Virginia & Ohio State & 17--6  & Virginia \\
304 & Maryland & Richmond   & 25--8  & Richmond \\
$\vdots$ &  &  &  & $\vdots$ \\
\bottomrule
\end{tabular}
\endgroup
\end{table}

\section{Additional considerations}

\subsubsection{On running up the score}

Landmark studies conducted by Barrow et al. and Annis and Craig show that score differential-based methods tend to be both more predictive and more likely to converge on an interpretable solution than those that use only win-loss data. (\cite{Annis2005}, \cite{Barrow2013a}) Regardless, many methods exclude score data, including Wesley Colley’s Colley Matrix. Colley justifies his use of win-loss ratios as a means to “keep it simple,” explaining that he didn’t want his method to require “ad hoc adjustments” to adjust for “runaway scores.” \cite{Colley2002}

Safe3Step isn't vulnerable to score inflating gamesmanship because it reduces the S3S point-value of a win against opponents that regularly get beat up on. The more a team runs up the score, the less that win against that opponent will be worth when calculating S3S points at the end of the season.\footnote{It's unlikely that a team would choose to run up the score (or minimize their score in a win) to adjust their or their opponent's S3S strength measurement to change the overall S3S point allocation table. Regardless, a simple "cap" on measured margins of victory at, say, +7 or -7 goals would solve such a problem.}

\subsubsection{Point spreads and gambling}
Algorithms based on margins of victory or "point spreads," share terminology with sports gambling. S3S, however, like most algorithms based on score data, fails to consider many factors that influence game outcomes, including personnel injuries, weather, morale, turf type, and others, limiting usefulness in predicting winners. S3S is intended to measure "deservedness" of an at-large pick—not to aid bettors.\footnote{These considerations are discussed in greater detail in the Powerwise whitepaper. \cite{FeldmanBomparola2025}}

\section{Conclusions}
Safe3Step was designed to aid NCAA Lacrosse committees with selecting which teams deserve at-large bids to their end-of-season championship tournaments. S3S's three-step system limits bias and disincentivizes gamesmanship by providing a "risk-reward" environment in which teams are: \\

\begin{list}{\Roman{myroman}.}{%
    \usecounter{myroman}%
    \setlength{\leftmargin}{1.5em}
    \setlength{\labelwidth}{1.5em}%
    \setlength{\labelsep}{0.5em}
    \setlength{\itemsep}{2pt}
    \setlength{\topsep}{0pt}
    \setlength{\parsep}{0pt}%
    \setlength{\partopsep}{0pt}%
}
  \item Rewarded for wins against tough opponents. 
  \item Fairly punished for losses.
  \item Never punished for a win, no matter how small.\\
\end{list}

In closing, S3S decidedly improves on the RPI in terms of objectivity, simplicity, and transparency. However, as mentioned, further analysis of its merits and pitfalls is left to future work. 

\onecolumn
\nocite{*}
\bibliographystyle{plainnat}
\bibliography{references}

\begin{thebibliography}{5}
\providecommand{\natexlab}[1]{#1}
\providecommand{\url}[1]{\texttt{#1}}
\expandafter\ifx\csname urlstyle\endcsname\relax
  \providecommand{\doi}[1]{doi: #1}\else
  \providecommand{\doi}{doi: \begingroup \urlstyle{rm}\Url}\fi

\bibitem[Annis and Craig(2005)]{Annis2005}
D.~H. Annis and B.~A. Craig.
\newblock Hybrid paired comparison analysis, with applications to the ranking of college football teams.
\newblock \emph{Journal of Quantitative Analysis in Sports}, 1\penalty0 (1), 2005.
\newblock \doi{10.2202/1559-0410.1000}.
\newblock URL \url{https://doi.org/10.2202/1559-0410.1000}.

\bibitem[Barrow et~al.(2013)Barrow, Drayer, Elliott, Gaut, and Osting]{Barrow2013a}
D.~Barrow, I.~Drayer, P.~Elliott, G.~Gaut, and B.~Osting.
\newblock Ranking rankings: An empirical comparison of the predictive power of sports ranking methods.
\newblock \emph{Journal of Quantitative Analysis in Sports}, 9\penalty0 (2):\penalty0 187--202, 2013.
\newblock \doi{10.1515/jqas-2013-0013}.
\newblock URL \url{https://doi.org/10.1515/jqas-2013-0013}.

\bibitem[Coleman et~al.(2010)Coleman, DuMond, and Lynch]{Coleman2010}
B.~J. Coleman, J.~M. DuMond, and A.~K. Lynch.
\newblock Evidence of bias in ncaa tournament selection and seeding.
\newblock \emph{Managerial and Decision Economics}, 31\penalty0 (7):\penalty0 431--452, 2010.

\bibitem[Colley(2002)]{Colley2002}
W.~N. Colley.
\newblock Colley’s bias free college football ranking method.
\newblock Technical report, Princeton University, 2002.
\newblock URL \url{https://arxiv.org/abs/cs/0208005}.

\bibitem[Feldman and Bomparola(2025)]{FeldmanBomparola2025}
L.~Feldman and M.~Bomparola.
\newblock Introducing powerwise (pwr): A pairwise and power rating method for selecting at-large teams to the ncaa division i men's lacrosse championship.
\newblock arXiv preprint arXiv:2508.04919, August 2025.
\newblock URL \url{https://arxiv.org/abs/2508.04919}.

\end{thebibliography}

\appendix

\newpage
\section*{Appendix A: Three Problems With the Ratings Percentage Index}
The Rating Percentage Index (RPI) statistic exhibits several flaws that affect its accuracy and reliability, especially for use in NCAA Lacrosse. Here are three major deficiencies illustrated by example cases:

\subsection*{Issue 1: Strength of Schedule (SOS) is Based on Opponent Performance, Not Team Performance}
The RPI’s SOS calculation focuses on the win percentage of a team’s opponents, ignoring it performed against those opponents. A hypothetical scenario: If Hampton (Men’s D1 lacrosse), ranked 76th in 2024, accepted an unlikely invite to the ACC and scheduled tougher competition, even if it lost all games by a wide margin, its RPI would jump, improving its rank by 38 places.

The RPI is more sensitive to a team's opponents’ performance than its own, distorting rankings. Its formula's reliance on opponent win percentages (OWP) and opponents’ opponents win percentages (OOWP) allows teams to artificially rise in the rankings merely by playing tougher competition, regardless of actual performance. To illustrate: Table 5 presents the actual RPI results from the 2023-2024 season and Table 6 presents the same results given our hypothetical scenario, showing Hampton’s RPI improvement in rating—from 0.3179 to 0.5053—and rank—from 76th to 38th—despite remaining the same team.

\begin{table}[H]
  \centering
  \small
  \caption{Hampton Dilemma: original vs.\ “ACC‑ified” RPI rankings}
  \begin{subtable}[t]{0.48\textwidth}
    \centering
    \caption{Original RPI Rankings}
    \label{tab:B1}
    \begin{tabular}{@{}r l c r r@{}}
      \toprule
      \textbf{RPI Rank} & \textbf{Team Name}   & \textbf{RPI} & \textbf{Wins} & \textbf{Losses} \\
      \midrule
      70 & Mt.\ St.\ Mary’s & 0.3663 &  1 & 14 \\
      71 & St.\ Bonaventure & 0.3538 &  1 & 11 \\
      72 & Wagner           & 0.3468 &  1 & 12 \\
      73 & Queens           & 0.3353 &  2 & 11 \\
      74 & Mass–Lowell      & 0.3350 &  0 & 12 \\
      75 & Lindenwood       & 0.3235 &  0 & 12 \\
      76 & \textbf{Hampton} & \textbf{0.3179} &  0 & 13 \\
      \bottomrule
    \end{tabular}
  \end{subtable}%
  \hfill
  \begin{subtable}[t]{0.48\textwidth}
    \centering
    \caption{Adjusted (ACC) Schedule RPI}
    \label{tab:B2}
    \begin{tabular}{@{}r l c r r@{}}
      \toprule
      \textbf{RPI Rank} & \textbf{Team Name}   & \textbf{RPI} & \textbf{Wins} & \textbf{Losses} \\
      \midrule
      38 & \textbf{Hampton} & \textbf{0.5053} &  0 & 13 \\
      39 & Vermont          & 0.4907          &  8 &  8 \\
      40 & Drexel           & 0.4883          &  5 &  9 \\
      41 & Air Force        & 0.4876          &  9 &  6 \\
      42 & Quinnipiac       & 0.4874          &  9 &  5 \\
      43 & Brown            & 0.4790          &  3 & 11 \\
      44 & LIU              & 0.4730          & 10 &  4 \\
      \bottomrule
    \end{tabular}
  \end{subtable}%
  \label{tab:hampton‑dilemma}
\end{table}

\noindent\textit{Note.} Hampton ranked 76th during the 2024 season. When its schedule is “ACC‑ified,” it jumps to 38th.

\subsection*{Issue 2: Ignoring Goal Differentials Invites Inaccuracy}
The second major issue is that the RPI only accounts for wins and losses, disregarding scores. Consider Team A (which defeats teams by large margins) versus Team B (which wins against the same teams by narrow margins). The RPI system would rate both teams similarly, despite Team A’s superior performance, overlooking the fact that a decisive win (e.g., 15-0) indicates a stronger team than a close win (e.g., 13-12). Ignoring score data reduces the granularity of analysis and leads to less accurate rankings.

\subsection*{Issue 3: Hypersensitivity to Irrelevant Games}
The RPI’s sensitivity to small changes in irrelevant or inconsequential games further compromises its accuracy. The example illustrated in Tables 6 and 7 shows how a remote upset, which has little bearing on tournament qualification, can significantly impact rankings. 

The left and right panels of Table 6 present RPI-based ranking lists based on the same data, except for a reversal in the result of a close game played early in the season between two relatively weak teams: Delaware and Lafayette. The right panel of Table 6 shows that the rankings of multiple top-20 teams have shifted despite Lafayette and Delaware having little direct interaction with the tournament contenders. Such hypersensitivity undermines the sense that teams control their own destiny.

\begin{table}[H]
  \centering
  \small
  \caption{RPI rankings before vs.\ after a single‑game upset\\
           (teams whose rank changed are in \textbf{bold})}
  \label{tab:B3-full}
  \begin{tabular}{@{}r l c l c@{}}
    \toprule
    & \multicolumn{2}{c}{\textbf{Original RPI}} 
    & \multicolumn{2}{c}{\textbf{Perturbed RPI}} \\ 
    \cmidrule(lr){2-3}\cmidrule(lr){4-5}
    \textbf{Rank} & \textbf{Team}        & \textbf{RPI} 
                  & \textbf{Team}        & \textbf{RPI} \\
    \midrule
     1 & Notre Dame        & 0.7100 & Notre Dame        & 0.7097 \\
     2 & Duke              & 0.6632 & Duke              & 0.6631 \\
     3 & Johns Hopkins     & 0.6520 & Johns Hopkins     & 0.6517 \\
     4 & Syracuse          & 0.6404 & Syracuse          & 0.6381 \\
     5 & Virginia          & 0.6371 & Virginia          & 0.6368 \\
     6 & Denver            & 0.6246 & Denver            & 0.6246 \\
     7 & Maryland          & 0.6223 & Maryland          & 0.6223 \\
     8 & \textbf{Princeton}         & 0.6158 & \textbf{Penn State} & 0.6157 \\
     9 & \textbf{Penn State}        & 0.6158 & \textbf{Princeton}  & 0.6157 \\
    10 & Georgetown        & 0.6147 & Georgetown        & 0.6146 \\
    11 & Penn              & 0.6044 & \textbf{Cornell}    & 0.6038 \\
    12 & \textbf{Cornell}           & 0.6041 & \textbf{Yale}       & 0.6018 \\
    13 & \textbf{Michigan}          & 0.6019 & \textbf{Penn}       & 0.6018 \\
    14 & \textbf{Yale}              & 0.6017 & \textbf{Michigan}   & 0.5994 \\
    15 & St.~Joseph’s      & 0.5970 & St.~Joseph’s      & 0.5967 \\
    \bottomrule
  \end{tabular}
\end{table}

Table 7 repeats this experiment with Power Ratings-based ranking lists. Only one, explainable (given knowledge about Army and Towson's record against Delaware and Lafayette), ranking change occurs in the right panel.

\vspace{1em}

\begin{table}[H]
  \centering
  \small
  \caption{Power Ratings before vs.\ after the same upset\\
           (teams whose rank changed are in \textbf{bold}}
  \label{tab:B4-full}
  \begin{tabular}{@{}r l c l c@{}}
    \toprule
    & \multicolumn{2}{c}{\textbf{Original PR}} 
    & \multicolumn{2}{c}{\textbf{Perturbed PR}} \\ 
    \cmidrule(lr){2-3}\cmidrule(lr){4-5}
     \textbf{Rank} 
    & \textbf{Team}        & \textbf{PR} 
    & \textbf{Team}        & \textbf{PR} \\
    \midrule
     1  & Notre Dame      & 99.90 & Notre Dame      & 99.90 \\
     2  & Duke            & 97.51 & Duke            & 97.52 \\
     3  & Virginia        & 97.38 & Virginia        & 97.37 \\
     4  & Syracuse        & 97.02 & Syracuse        & 97.00 \\
     5  & Penn State      & 96.90 & Penn State      & 96.91 \\
     6  & Johns Hopkins   & 96.77 & Johns Hopkins   & 96.77 \\
     7  & Princeton       & 96.46 & Princeton       & 96.47 \\
     8  & Georgetown      & 95.78 & Georgetown      & 95.78 \\
     9  & Maryland        & 95.67 & Maryland        & 95.68 \\
    10  & Cornell         & 95.67 & Cornell         & 95.68 \\
    11  & Yale            & 95.47 & Yale            & 95.49 \\
    12  & Denver          & 95.47 & Denver          & 95.47 \\
    13  & Michigan        & 95.25 & Michigan        & 95.21 \\
    14  & \textbf{Towson} & 94.80 & \textbf{Army}   & 94.82 \\
    15  & \textbf{Army}   & 94.68 & \textbf{Towson} & 94.69 \\
    \bottomrule
  \end{tabular}
\end{table}

\textbf{To conclude: }The problems with the RPI compromise its effectiveness as a ranking system, resulting in a failure to accurately reflect team ability and/or performance, undermining the sense that teams control their destiny on the field. This evidence highlights the need for accurate systems (like Power Ratings) that can incorporate additional data, such as goal differentials, and reduce the impact of arbitrary results. 
\end{document}